\documentclass[graybox]{article}

\usepackage{mathptmx}       
\usepackage{helvet}         
\usepackage{courier}        
\usepackage{type1cm}        
\usepackage{makeidx}         
\usepackage{graphicx}        
\usepackage{multicol}        
\usepackage[bottom]{footmisc}

\usepackage{caption,subfig}    
\usepackage{amsmath,amsthm,amssymb,amstext,esvect}  

\makeindex             

\begin{document}

\title{An Agent Based Modeling of Spatially Inhomogeneous Host-Vector Disease Transmission}

\author{Isti Rodiah, Wolfgang Bock, and Torben Fattler \\ TU Kaiserslautern \\ Technomathematics Group, Department of Mathematics \\ P.O. Box 3049, D-67653 Kaiserslautern \\ rodiah@mathematik@uni-kl.de, \\ bock@mathematik.uni-kl.de,\\ fatller@mathematik.uni-kl.de}

\date{}

\maketitle

\abstract{In this article we consider a microscopic model for host-vector disease transmission based on configuration space analysis. Using Vlasov scaling we obtain the corresponding mesoscopic (kinetic) equations, describing the density of susceptible and infected compartments in space. The resulting system of equations can be seen as a generalization to a spatial SISUV model.}

\section{Introduction}
Kinetic models in disease spread are often a starting point of theoretical studies in epidemiology. In disease spread dynamics, it is in many cases known how the infection dynamics evolves on the level of particle interaction. The modeling of spatial disease spread from a microscopic agent-to-agent model has been already done e.g.~in a cancer model by \cite{FFHKKK15} and in \cite{BFRT17a, BFRT17b} for a direct contact disease transmission. PDE models for a spatial dynamic of SISUV dynamics have been proposed see e.g.~\cite{RD13, RASS12}, however, up to the authors knowledge, there has not been shown yet, that these PDE models are well-defined Vlasov scaling limits arising from a particle system model on the agent-to-agent-interaction, hence the microscopic level.   

Disease transmission represents the contact between host and vector in host-vector diseases. A series of different models for vector-bourne diseases such as Dengue fever including stochastic and deterministic models have been proposed, see e.g.~\cite{RMSAS16} and references therein. The models described above do not provide any information about the spatial spread of a disease. In the SIR (Susceptible-Infected-Recovered) model case, an advection-diffusion equation has been identified as the limiting equation in a long time scaling limit, see \cite{CS11}. Another approach in incorporating spatial information for the SIR model may also be found in \cite{STW16} and recently \cite{BJ18}. On the macroscopic level, the models are very flexible for describing the different aspects of disease dynamics. However, for many different diseases the infection mechanism is only known on the microscopic, i.e.,~agent-to-agent level.
To consider both microscopical modeling and spatial resolution, we describe the disease dynamics by means of an interacting particle system with suitable interaction potentials. Fundamental in this area are dynamics in so-called marked configuration spaces \cite{FKO13}. These techniques together with a proper scaling of the microscopic system, the so-called Vlasov scaling, have been recently used to model the dynamics of cancer cells \cite{FFHKKK15}.

\section{Microscopic Model}
In \cite{R18} the host-vector disease transmission is modeled via marked configuration spaces. 
The \emph{configuration space $\Gamma$ over $\mathbb{R}^2$}  is defined by  
$$\Gamma := \Gamma_{\mathbb{R}^2} := \big\{ \gamma \subset \mathbb{R}^2 \, \big| \, \# (\gamma \cap K) < \infty \mbox{ for all } K \subset \mathbb{R}^2 \mbox{ compact} \big \},$$
where $\# A$ denotes the cardinality of a set $A$.
Given four copies of the space $\Gamma$, denoted by $\Gamma^S$, $\Gamma^I$, $\Gamma^U$, and $\Gamma^V$, let
$$ \Gamma^{4} := \big\{ \vv{\gamma} := (\gamma^{S}, \gamma^{I}, \gamma^{U}, \gamma^{V}) \in \Gamma^{S} \times \Gamma^{I} \times \Gamma^{U} \times \Gamma^{V} \, \big| \, \gamma^{i} \cap \gamma^{j} = \varnothing, \ i \neq j \big\}.$$
The model hence consists of four compartments gives as susceptible hosts (S), infected hosts (I), susceptible vectors (U), and infected vectors (V). We set up the model in the evolution of the aforementioned four-component system in the state space $\Gamma^{4}$.

\begin{figure}[htbp]
\centering
\includegraphics[scale=0.2]{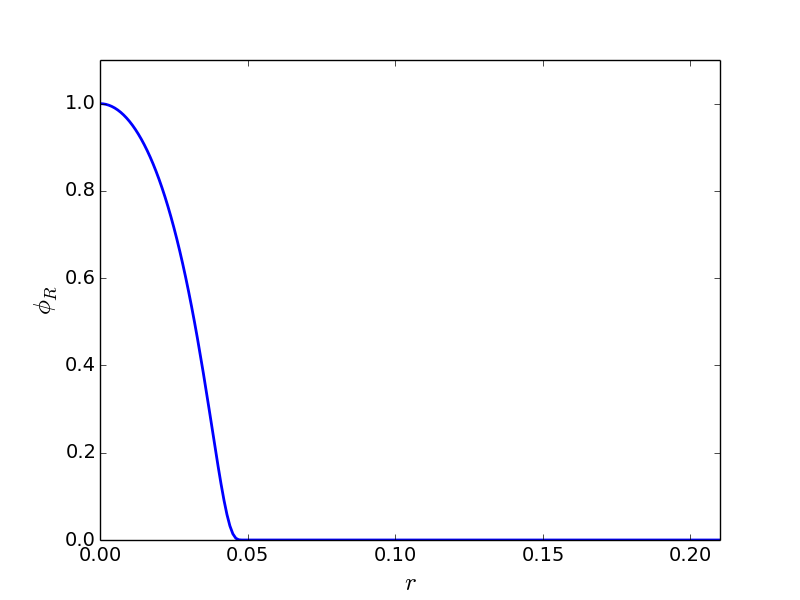}
\vspace{-0.5\baselineskip}
\caption{\footnotesize{The potential with $R = 0.05$}}
\label{fig.1} 
\end{figure}

For a specification of the infection rates, we use a potential depending on individual to individual distance
$$ [0,\infty)\ni r\mapsto \phi_{R}(r) :=\phi(r) \in [0,\infty), $$
with $R \in (0,\infty)$. One example of the potential is shown in Figure \ref{fig.1}. 
We set $\beta_{h} \in [0, 1]$ to be the risk of infection for a susceptible host to be in direct contact with an infected vector and $\beta_{v} \in [0, 1]$ to be the risk of infection for a susceptible vector to be in direct contact with an infected host. For fixed $x \in \gamma^{S}$, the infection rate for a single susceptible host located at $x$ in the surrounding $\gamma^{V} \in \Gamma^{V}$ is given by $c_{h} (x, \gamma^{V})$. For fixed $\tilde{x} \in \gamma^{U}$, the infection rate for a single susceptible vector located at $\tilde{x}$ in the surrounding $\gamma^{I} \in \Gamma^{I}$ is given by $c_{v} (\tilde{x}, \gamma^{I})$.
$$c_{h} (x, \gamma^{V}) = \beta_{h} \sum_{\tilde{y} \in \gamma^{V}} \phi \! \left( |x-\tilde{y}| \right), \qquad c_{v} (\tilde{x}, \gamma^{I}) = \beta_{v} \sum_{y \in \gamma^{I}} \phi \! \left( |\tilde{x}-y| \right).$$

In host-vector disease transmission, we define a couple of generators $L_{h}$ and $L_{v}$, that are the generators for host and vector, respectively. The disease dynamics are given by 
\begin{align} \label{eq.1}
\nonumber
(L_{h}F) \! \left( \vv{\gamma}\right) := & \sum_{x \in \gamma^{S}} c_{h} \! \left( x, \gamma^{V} \right) \Big( F \! \left(\gamma^{S} \setminus \{x\}, \gamma^{I} \cup \{x\}, \gamma^{U}, \gamma^{V} \right) - F \! \left(\vv{\gamma} \right) \Big) \\
 & + \sum_{y \in \gamma^{I}} \alpha_{h} \, \Big( F \! \left( \gamma^{S} \cup \{y\}, \gamma^{I} \setminus \{y\}, \gamma^{U}, \gamma^{V} \right) - F \! \left( \vv{\gamma} \right) \Big)
 \end{align}
 and 
 \begin{align} \label{eq.2}
(L_{v}F) \! \left(\vv{\gamma}\right)  :=  \sum_{\tilde{x} \in \gamma^{U}} c_{v} \! \left( \tilde{x}, \gamma^{I} \right) \Big( F \! \left(\gamma^{S}, \gamma^{I}, \gamma^{U} \setminus \{\tilde{x}\}, \gamma^{V} \cup \{\tilde{x}\} \right) - F \! \left(\vv{\gamma}\right) \Big),
\end{align}
where the function $c_{h}(x, \gamma^{V})$ is the infection rate of host, $\alpha_{h} \in [0, 1]$ is the constant recovery rate of host, and the function $c_{v}(\tilde{x}, \gamma^{I})$ is the infection rate of vector.

\section{Numerical Simulation}
In this section, we give a brief introduction into the numerical method which is used for simulation. The spread of the disease is modeled via a flip according to an infection rate, given via the Markov generator in Section 2. Since the infected individuals influence the infection rate at a certain point in the area, the computation of these rates is the main task. Briefly, the procedure of numerical implementation is as follows:
\begin{enumerate}
\item[(i)] Generate the state of individuals and distribute the individuals uniformly in space.
\item[(ii)] Calculate the transition rate or probability for each individual.
\item[(iii)] Generate random variable, then compare it with the transition rate. If the random variable is smaller than the transition rate, the state of the individual is changed. 
\end{enumerate}

\begin{figure}[htbp]
\captionsetup[subfloat]{farskip=0.pt, captionskip=-1.pt}
\centering
\subfloat[\footnotesize{$t = 0$}]{
  \includegraphics[width=29mm]{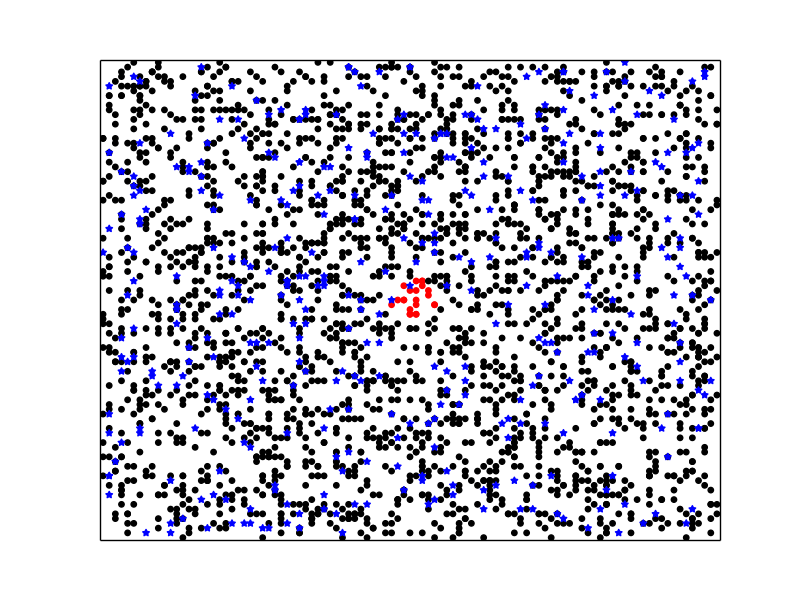}
}
\subfloat[\footnotesize{$t = 2$}]{
  \includegraphics[width=29mm]{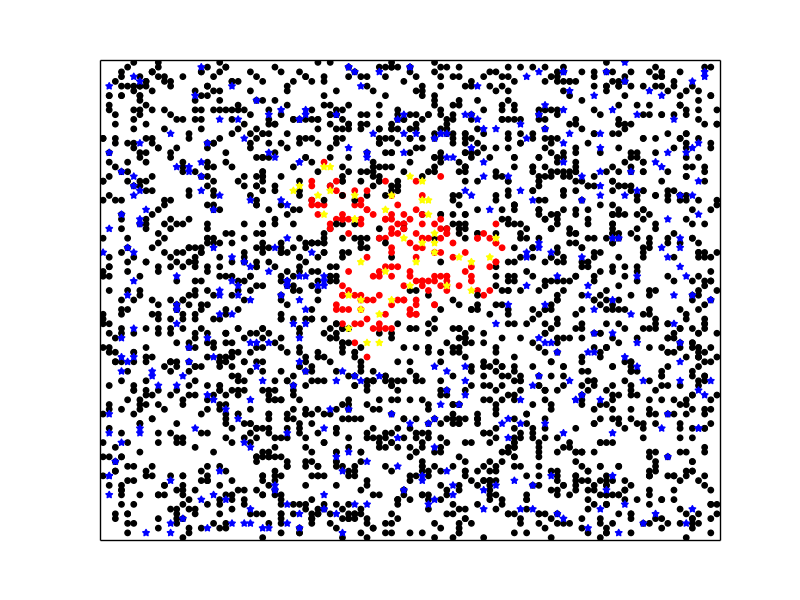}
}
\subfloat[\footnotesize{$t = 5$}]{
  \includegraphics[width=29mm]{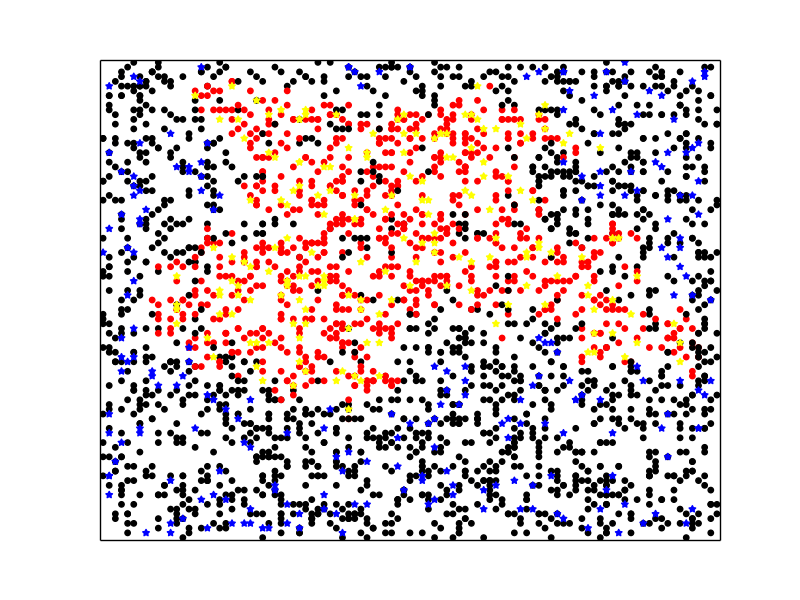}
}
\subfloat[\footnotesize{$t = 10$}]{
  \includegraphics[width=29mm]{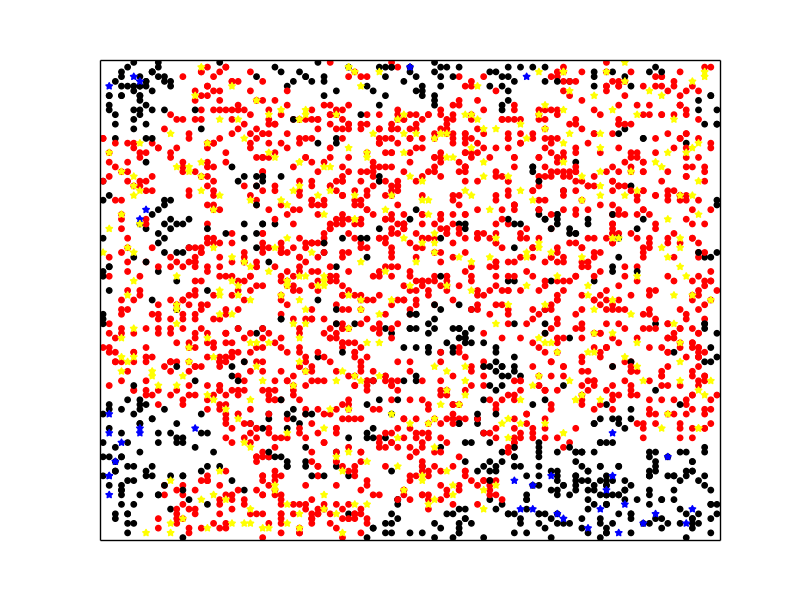}
}
\vspace{-0.5\baselineskip}
\caption{\footnotesize{SISUV model of 2500 host and 400 vectors included infection and recovery}}
\label{fig.2}
\end{figure}

We consider the area $[0,1] \times [0,1] \subset \mathbb{R}^2$ with $\beta_{h} = 0.1$, $\beta_{v} = 0.2$, and $\alpha_{h} = 0.14$. Figure~\ref{fig.2} shows a spatial distribution of hosts and vectors evolving in time. In this particular case, we have the initial number of infectious hosts $I(0) = 20$, susceptible hosts $S(0) = 2480$, infectious vectors $V(0) = 0$, and susceptible vectors $U(0) = 400$. Susceptible hosts are depicted as black spots, infected host as red spots,  susceptible vectors as blue spots and infected vectors as yellow spots.
\vspace{-3mm}

\subsection{Infection and Recovery}
\subsubsection{Comparison of Particle and Deterministic SISUV Model} 
In the particle model, we consider specific infection rates depending on the surrounding. However just considering the number of incidents should lead to an ODE system for a high number of particles. On the other hand, the standard ODE model assumes a uniform distribution of all particles from the beginning and neglects all behavior coming from spatial effects. Here, we compare the SISUV ODE system with the particle system for spatial uniformly distributed particles for $\phi = 1$ and $\phi=\phi_R$ from the previous section (the infection is localized). 
 
\begin{figure}[htbp]
\captionsetup[subfloat]{farskip=0.pt, captionskip=-1.pt}
\centering
\subfloat[\footnotesize{Host}]{
  \includegraphics[scale=0.2]{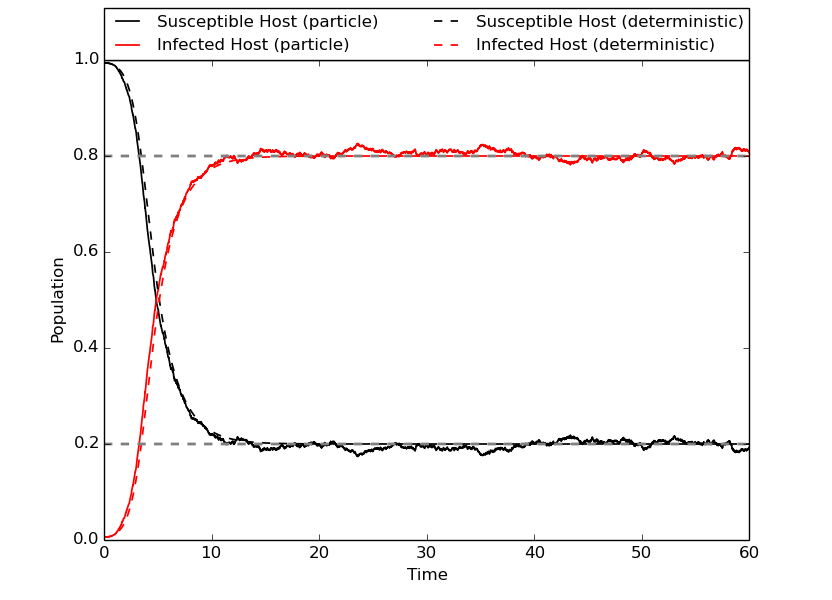}
}
\subfloat[\footnotesize{Vector}]{
  \includegraphics[scale=0.2]{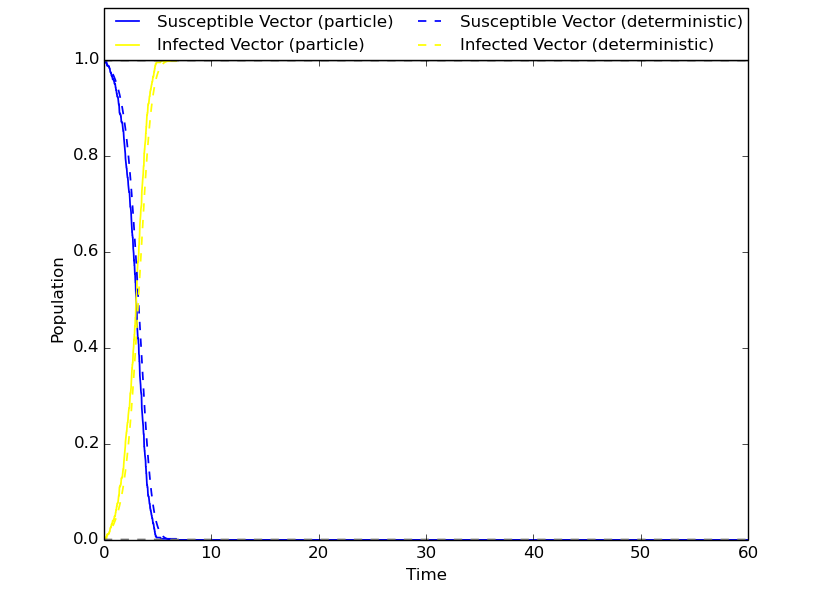}
}
\vspace{-0.5\baselineskip}
\caption{\footnotesize{The deterministic model and the particle model of 2500 host and 400 vectors with $\phi = 1$}}
\label{fig.3}
\end{figure}

We choose $\beta_{h} = 0.001$, $\beta_{v} = 0.002$, and $\alpha_{h} = 0.1$. First, we consider a constant potential of infection $\phi = 1$, i.e.˜ a susceptible host (vector) interacts with all infected vectors (hosts) via the same rate of infection. Figure \ref{fig.3} shows that the particle model is in good agreement to the classical SISUV model given by the ODE system
$$\frac{d}{dt}S(t) = -\beta_{h} \, S(t) \, V(t) + \alpha_{h} \, I(t), \qquad \frac{d}{dt}I(t) = \beta_{h} \, S(t) \, V(t) - \alpha_{h} \, I(t),$$
$$\frac{d}{dt}U(t) = -\beta_{v} \, U(t) \, I(t), \qquad \qquad \frac{d}{dt}V(t) = \beta_{v} \, U(t) \, I(t), $$
where in this case $\beta_{h}=0.001$, $\beta_{v}=0.002$, and $\alpha_{h}=0.1$.

\begin{figure}[htbp]
\captionsetup[subfloat]{farskip=0.pt, captionskip=-1.pt}
\centering
\subfloat[\footnotesize{Host}]{
  \includegraphics[scale=0.2]{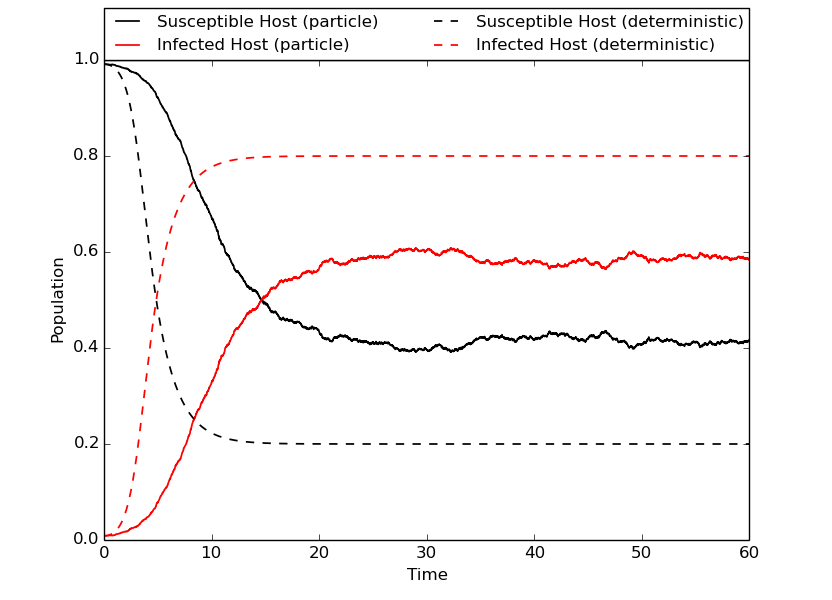}
}
\subfloat[\footnotesize{Vector}]{
  \includegraphics[scale=0.2]{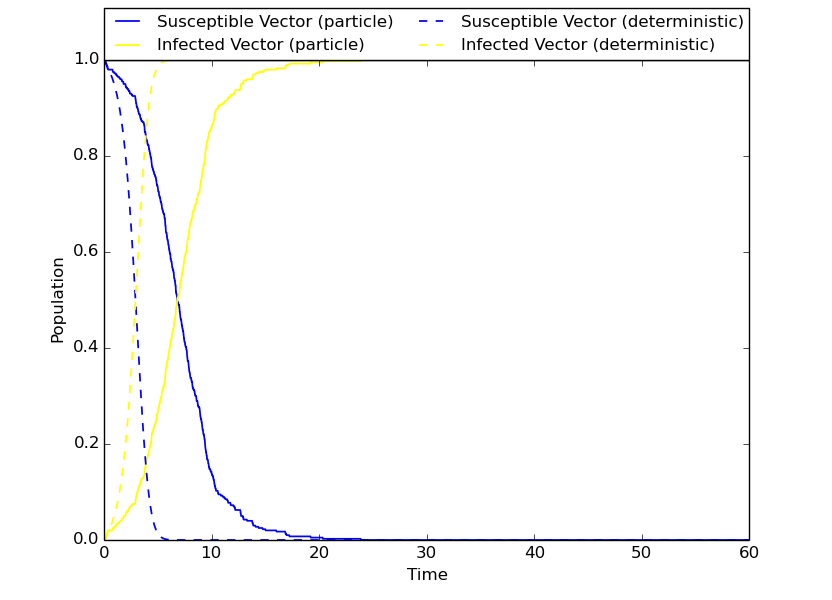}
}
\vspace{-0.5\baselineskip}
\caption{\footnotesize{The deterministic model and the particle model of 2500 host and 400 vectors with $\phi_{R}$}}
\label{fig.4}
\end{figure}

Then, we consider the particle model with a potential of infection $\phi_R$ i.e., infections are possible just if susceptible host (vector) and infectious vectors (hosts) are sufficiently close to each other. Figure \ref{fig.4} shows that the dynamics of the particle model is "slower'' than the classical SISUV model. This is due to the fact that individuals just interact locally in the particle model, while they are assumed to interact globally in the classical SISUV model. In both cases, the different simulations in Figure \ref{fig.3} and \ref{fig.4} show the same qualitative behavior, although having different infection radius leads to different infection rates.

\subsubsection{Comparison of Particle Model and Kinetic Equation}
\begin{figure}[htbp]
\captionsetup[subfloat]{farskip=0.pt, captionskip=-1.pt}
\centering
\subfloat[\footnotesize{$t = 0$}]{
  \includegraphics[width=29mm]{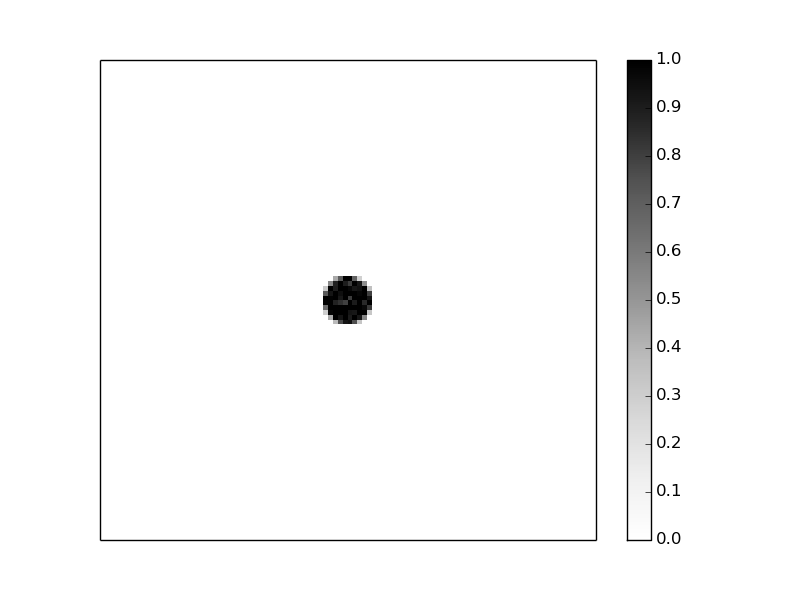}
}
\subfloat[\footnotesize{$t = 0.5$}]{
  \includegraphics[width=29mm]{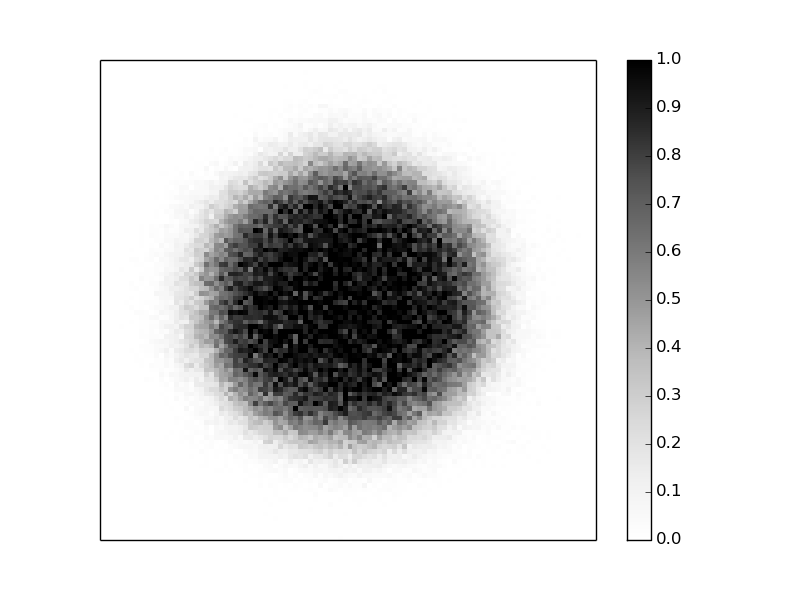}
}
\subfloat[\footnotesize{$t = 1$}]{
  \includegraphics[width=29mm]{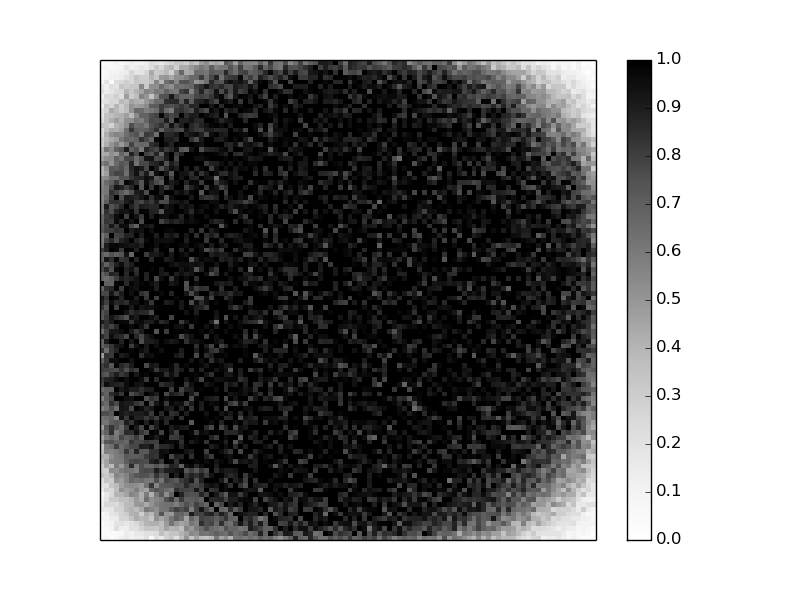}
}
\subfloat[\footnotesize{$t = 2$}]{
  \includegraphics[width=29mm]{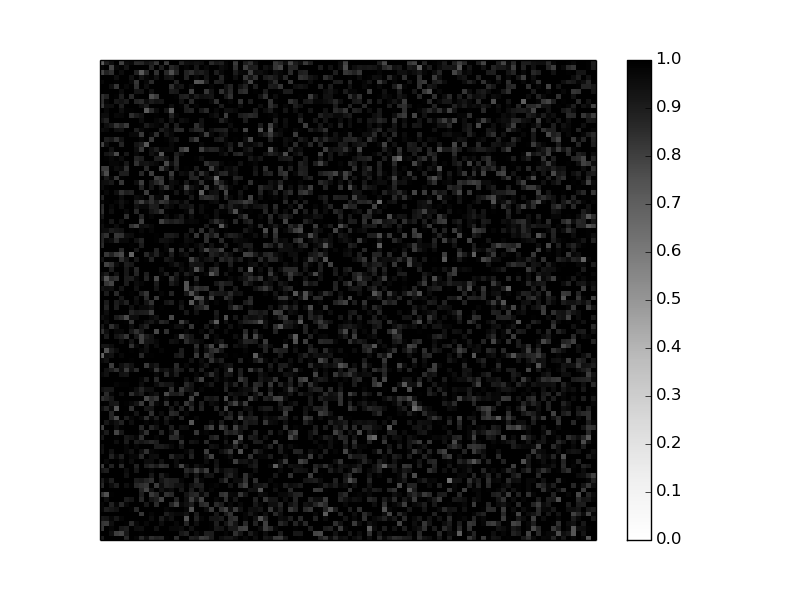}
}
\vspace{-0.5\baselineskip}
\caption{\footnotesize{Average of a hundred runs of the particle model of 10000 hosts and 900 vectors for the infected state of host}}
\label{fig.5}
\end{figure}

\begin{figure}[htbp]
\captionsetup[subfloat]{farskip=0.pt, captionskip=-1.pt}
\centering
\subfloat[\footnotesize{$t = 0$}]{
  \includegraphics[width=29mm]{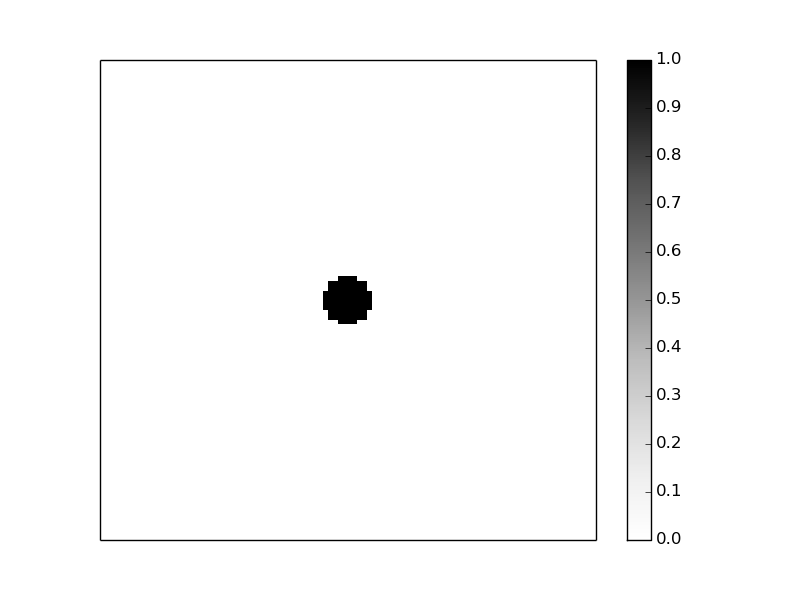}
}
\subfloat[\footnotesize{$t = 0.5$}]{
  \includegraphics[width=29mm]{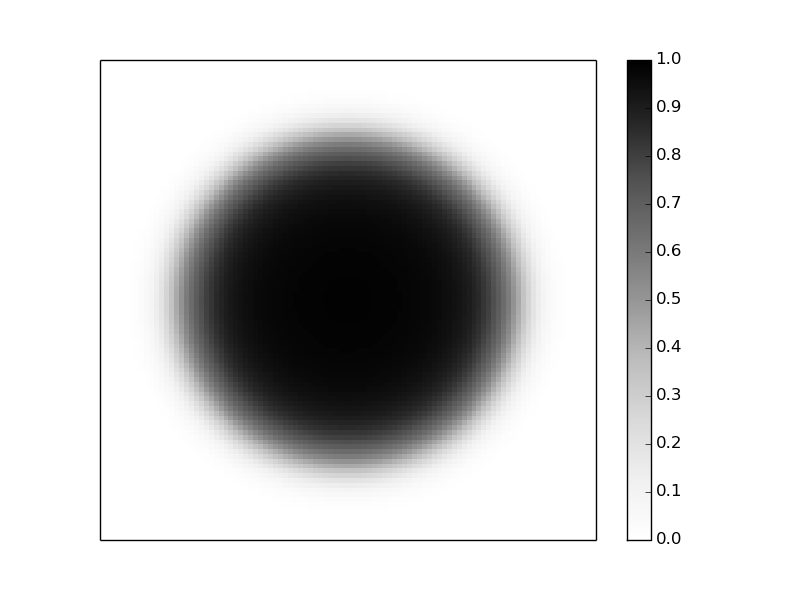}
}
\subfloat[\footnotesize{$t = 1$}]{
  \includegraphics[width=29mm]{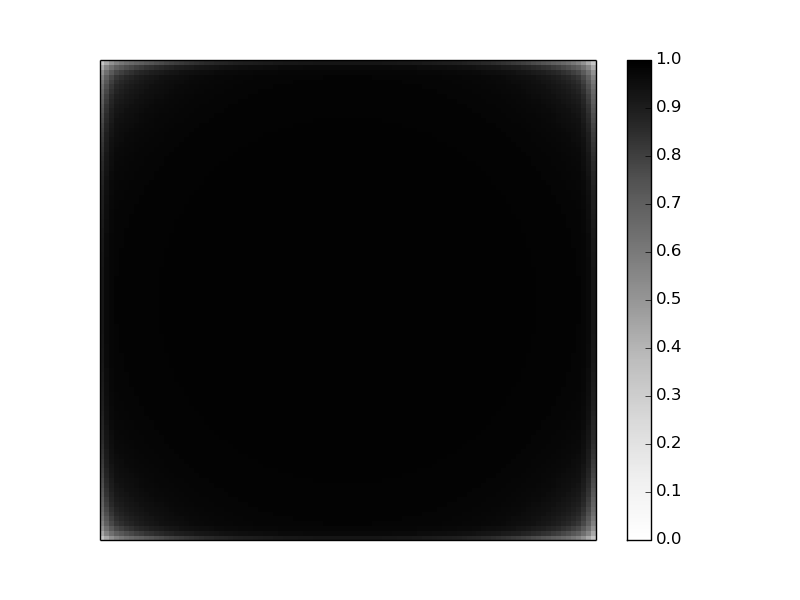}
}
\subfloat[\footnotesize{$t = 2$}]{
  \includegraphics[width=29mm]{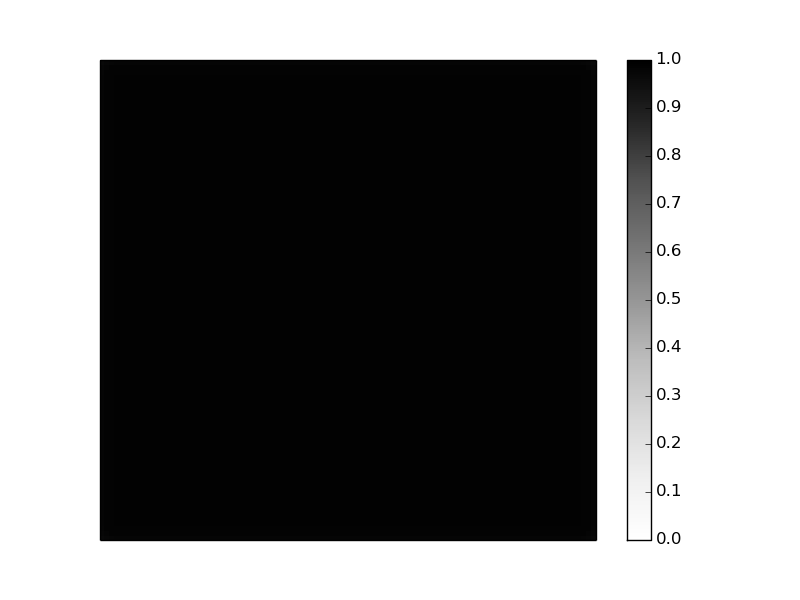}
}
\vspace{-0.5\baselineskip}
\caption{\footnotesize{Numerical solution of the kinetic equation for the infected state of host}}
\label{fig.6}
\end{figure}

\begin{figure}[htbp]
\captionsetup[subfloat]{farskip=0.pt, captionskip=-1.pt}
\centering
\subfloat[\footnotesize{$t = 0$}]{
  \includegraphics[width=29mm]{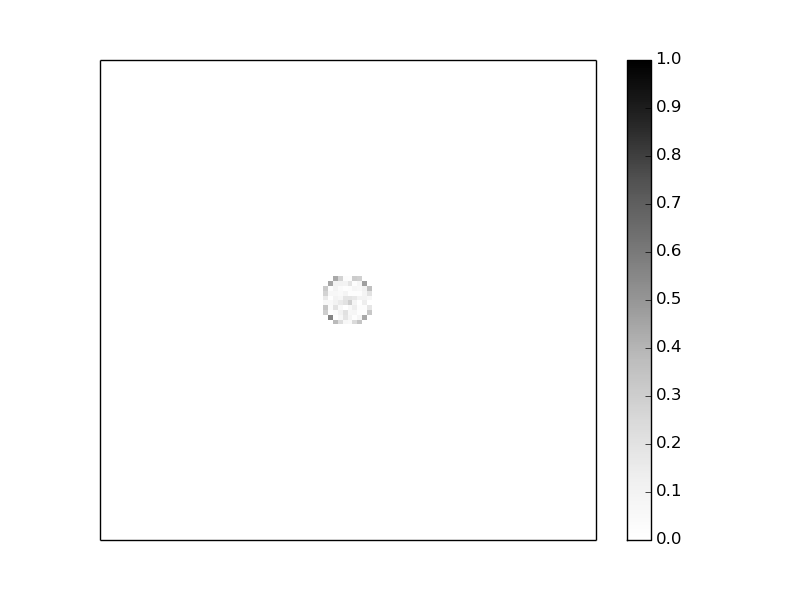}
}
\subfloat[\footnotesize{$t = 0.5$}]{
  \includegraphics[width=29mm]{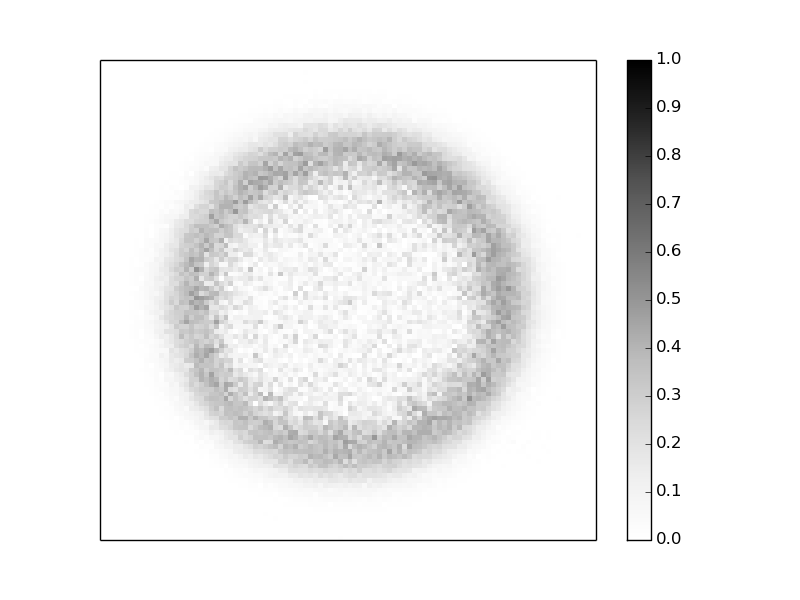}
}
\subfloat[\footnotesize{$t = 1$}]{
  \includegraphics[width=29mm]{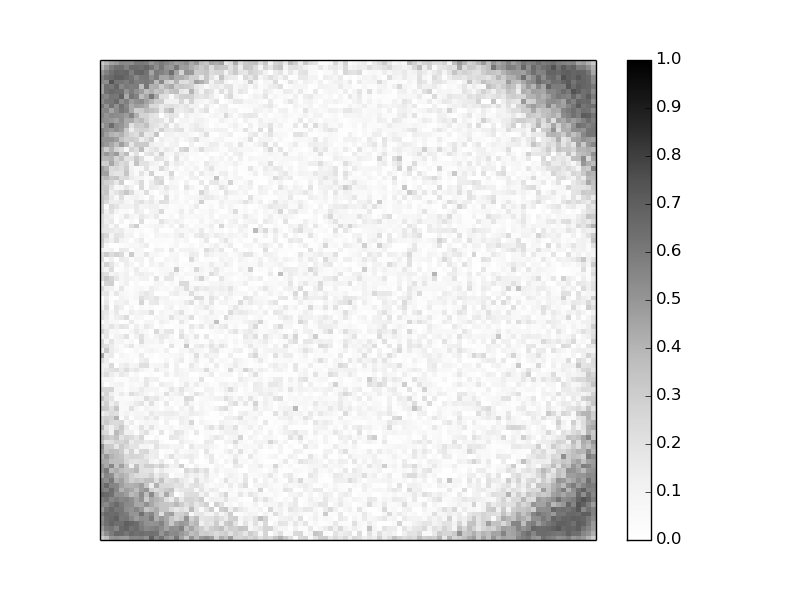}
}
\subfloat[\footnotesize{$t = 2$}]{
  \includegraphics[width=29mm]{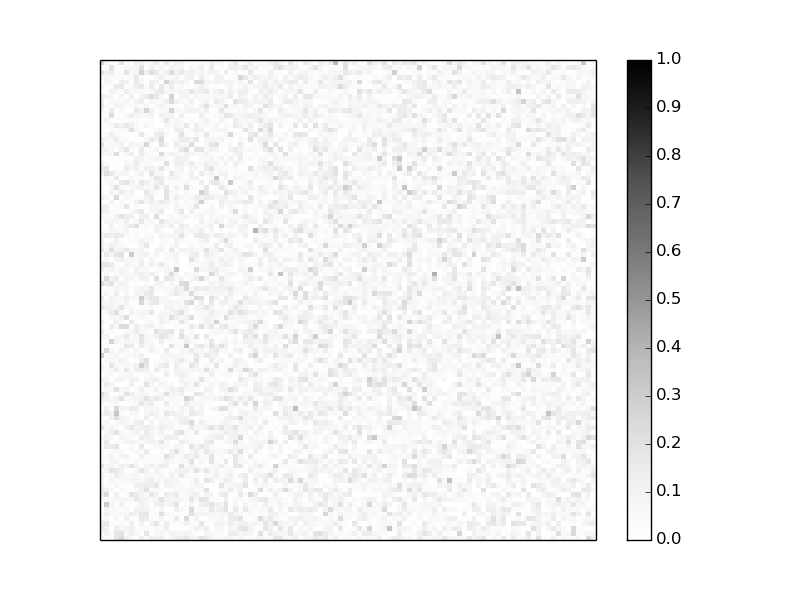}
}
\vspace{-0.5\baselineskip}
\caption{\footnotesize{Difference between the average of the particle model and the kinetic system for the infected state of host in each sub-domians}}
\label{fig.7}
\end{figure}

\begin{figure}[htbp]
\captionsetup[subfloat]{farskip=0.pt, captionskip=-1.pt}
\centering
\subfloat[\footnotesize{Host}]{
  \includegraphics[scale=0.2]{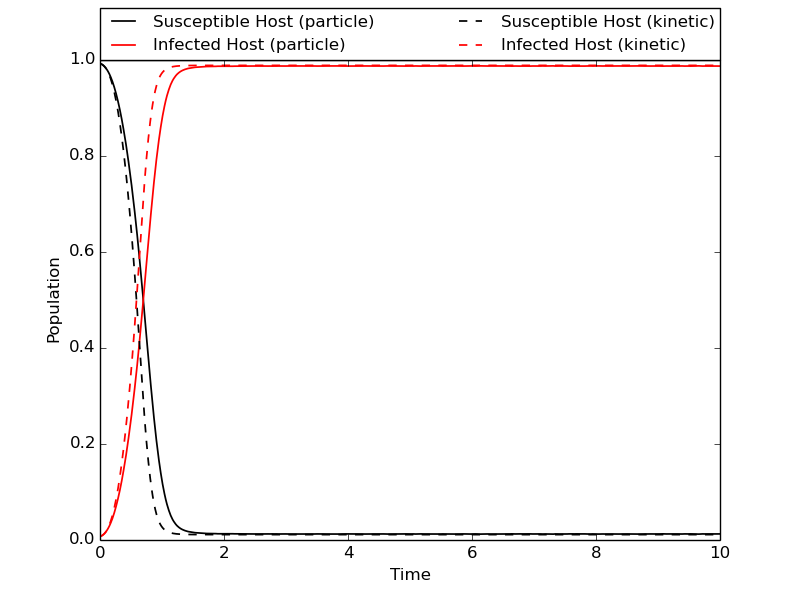}
}
\subfloat[\footnotesize{Vector}]{
  \includegraphics[scale=0.2]{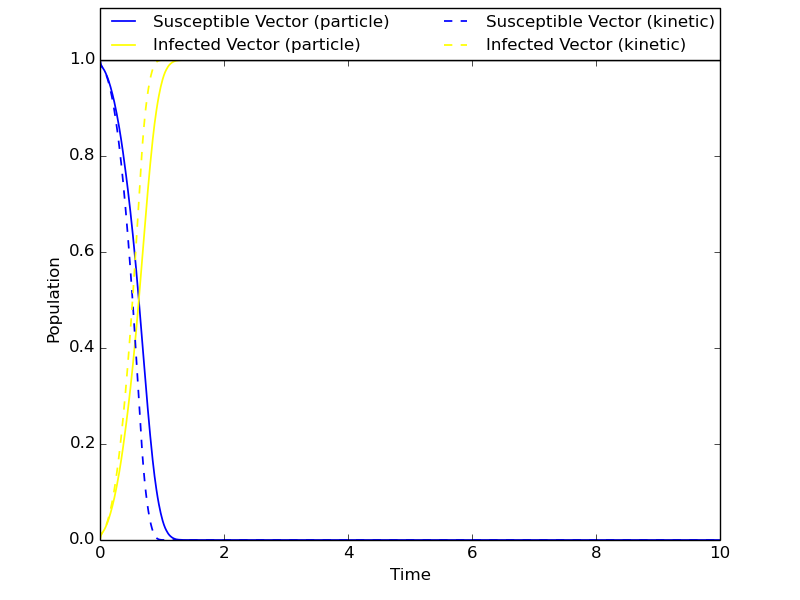}
}
\vspace{-0.5\baselineskip}
\caption{\footnotesize{Average of a hundred runs of the particle model and the kinetic system}}
\label{fig.8}
\end{figure}

We compare the kinetic equation with averaged runs of the particle simulation. For this purpose, we choose $\beta_{h} = 0.1$, $\beta_{v} = 0.2$, $\alpha_{h} = 0.14$,  $N = 10000$, and $M = 900$. For the kinetic system, we consider an equidistant spatial distribution of particles. For the simulation we use an initially random uniform spatial distribution of particles. Figure \ref{fig.5} shows the spatial distribution of averaged runs of the particle model. We partition the domain $[0,1] \times [0,1]$ in $10000$ sub-domains. The kinetic equation is solved via a standard finite differences method with $\Delta x = \frac{1}{100}$ and $\Delta t = 0.01$. Figure \ref{fig.6} shows the spatial solution of the kinetic equations. Figure \ref{fig.7} shows the difference between the average of the particle model and the kinetic equation in each sub-domains. The comparison between the dynamics in the kinetic and the particle approximation is shown in Figure \ref{fig.8}.

\end{document}